\documentclass[apj]{emulateapj}
\begin{document}
\shorttitle{Spectroscopic Confirmation of Pisces}
\shortauthors{Kollmeier et al.}
\newcommand{\msun}{M_{\odot}}
\newcommand{\kms}{\, {\rm km\, s}^{-1}}
\newcommand{\cm}{\, {\rm cm}}
\newcommand{\gm}{\, {\rm g}}
\newcommand{\erg}{\, {\rm erg}}
\newcommand{\kel}{\, {\rm K}}
\newcommand{\kpc}{\, {\rm kpc}}
\newcommand{\mpc}{\, {\rm Mpc}}
\newcommand{\seg}{\, {\rm s}}
\newcommand{\kev}{\, {\rm keV}}
\newcommand{\hz}{\, {\rm Hz}}
\newcommand{\etal}{et al.\ }
\newcommand{\yr}{\, {\rm yr}}
\newcommand{\mpyr}{{\rm mas}\, {\rm yr}^{-1}}
\newcommand{\gyr}{\, {\rm Gyr}}
\newcommand{\eq}{eq.\ }
\def\arcsec{''\hskip-3pt .}
\title{Spectroscopic Confirmation of the Pisces Overdensity\altaffilmark{1}} 
\author{Juna A. Kollmeier\altaffilmark{2},
Andrew Gould\altaffilmark{3}, 
Stephen Shectman\altaffilmark{2},
Ian B. Thompson\altaffilmark{2}, 
George W. Preston\altaffilmark{2}, 
Joshua D. Simon\altaffilmark{2},
Jeffrey D. Crane\altaffilmark{2}, \v{Z}eljko Ivezi\'{c}\altaffilmark{4}, \& Branimir Sesar\altaffilmark{4} 
}
\altaffiltext{1}{This paper includes data gathered with the 6.5 meter Magellan Telescopes located at Las Campanas Observatory,
Chile.}
\altaffiltext{2}{Observatories of the Carnegie Institution of Washington,
  813 Santa Barbara Street, Pasadena, CA 91101}
\altaffiltext{3}{Department of Astronomy, The Ohio State University,
  4051 McPherson Laboratory, Columbus, OH, 43210}
\altaffiltext{4}{Department of Astromony, University of Washington, Box 351580, Seattle, WA 98195-1580}

\begin{abstract}
  We present spectroscopic confirmation of the ``Pisces Overdensity'', also known as ``Structure J'', a photometric
  overdensity of RR Lyrae stars discovered by the Sloan Digital Sky
  Survey (SDSS) at an estimated photometric distance of $\sim 85$kpc.  We measure radial velocities for 8 RR Lyrae stars within Pisces.  We find that 5 of the 8 stars have heliocentric
  radial velocities within a narrow range of $-87\,{\rm
    km\,s^{-1}}< v_r < -67\,{\rm km\,s^{-1}}$, suggesting that the
  photometric overdensity is mainly due to a physically associated
  system, probably a dwarf galaxy or a disrupted galaxy.  Two of the remaining 3 stars differ from one another by only 9 km/s, but it
  would be premature to identify them as a second system.
\end{abstract}

\section{INTRODUCTION}

In the currently favored picture of galactic structure formation, the
Milky Way had a tumultuous early history.  Continuously bombarded
from an early age by other galaxies big and small, the Milky Way has
been roiled by mergers --- a picture first put forth by \citet{sz78}.
The fossil record of the Milky Way's history can be seen today in the
debris left over from this cosmic carnage -- the stellar streams that
appear to be the shredded remains of galaxies past.  The SDSS has had
a major impact in breaking open this field.  A variety of photometric
techniques have been developed to search through the large photometric
database of the SDSS to uncover these systems (e.g. Newberg et
al. 2002, Willman et al. 2005, Belokurov et al. 2006, Grillmair et
al. 2006).  The primary search algorithms rely on a combination of
correlations in position on the sky and position in the
color-magnitude diagram (CMD).  To date, a significant number of such
structures have been located.  However, the census of structures is
still very far from complete, particularly at large Galactocentric
distances.  The key issue is sensitivity: at very low intrinsic
luminosities these objects can only be disentangled from foreground
stars interior to $\sim$50kpc and SDSS star counts are only sensitive
to surface brightnesses above $\sim 30$ mag/arcsec$^{-2}$.  A novel
technique, first recognized by \citet{kinman63}, to push the limits
of finding substructure within the Milky Way is to make use of the
photometric variability of RR Lyrae stars.

The repeated scans of SDSS Stripe-82 have allowed the discovery of Milky Way
substructures via their populations of RR Lyrae stars, which are standard candles, can be seen to
large distances, and show distinctive light curves.  The structure
discussed in this {\it Letter} was first identified as an overdensity
of RR Lyrae stars by Sesar et al. (2007) as part of a larger study of
all RR Lyrae stars in Stripe-82. They termed this overdensity
``Structure J'', one among several such structures located.
In a subsequent work, Watkins et al. (2009) independently found what
appears to be the same structure, which they termed the ``Pisces
Overdensity''.  The photometric identification of these overdensities
is a crucial first step in locating more streams and galaxies interior
to the virial radius of the Milky Way.  However, photometry provides
two phase-space dimensions very accurately (angular position) and a
third only very crudely (distance).  In order to show definitively
whether photometric overdensities are, in fact, truly part of a common
structure as opposed to a chance concentration, it is necessary to
obtain data for additional phase-space dimension(s), which could in principle
be either radial velocities or proper motions \citep[e.g.,][]{simon07}.  In this {\it Letter},
we report on the first results of a campaign to confirm distant
structures in the Milky Way halo as defined by RR Lyrae and giant stars.

\section{Observations}

\subsection{Target Selection}
\begin{figure*}
\plotone{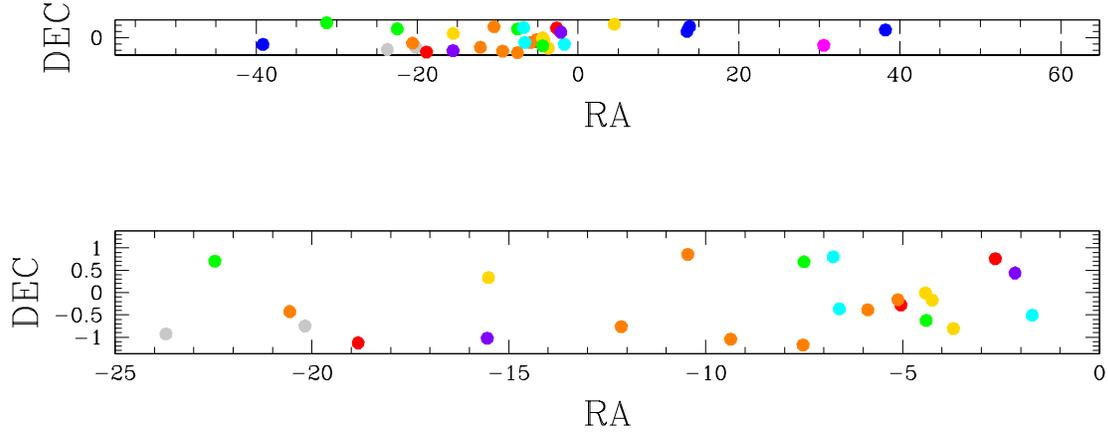}
\caption{RR Lyrae stars in Stripe-82.  {\bf Top Panel}: positions of RR Lyrae stars with magnitudes $19.9<V_0<20.8$
in the SDSS Stripe-82, color coded in 0.1 mag intervals, brightest to faintest:
(gray, red, orange, yellow, green, cyan, blue, magenta, purple). Notice the extreme concentration at RA $-25^\circ$ --
 $0^\circ$. {\bf Bottom Panel}:  Zoom of Panel A in the region of the concentration.}
\label{fig:structurej}
\end{figure*}

In Figure~\ref{fig:structurej} we show all of the identified RR Lyrae
stars in the apparent magnitude range $19.9\le V_0 \le 20.8$ that lie
in SDSS Stripe-82, which covers a $3^\circ \times 118^\circ$ stripe
along the celestial equator.  The positions and $V_0$ magnitudes of
these objects come from the latest Stripe-82 RR Lyrae catalog \citep{sesar09}.  There is a very clear overdensity in the RA
interval 335--360.  With only about 30\% of the stripe area, this
sub-region contains 24 of the 31 RR Lyrae stars (77\%).  For our first
campaign we focused on the concentration located at RA$\sim
355^{\circ}$ and selected 8 objects near this overdensity based on
their close 2D proximity and their similar median magnitudes.  

\subsection{Light Curve Analysis}
\begin{figure*}
\plotone{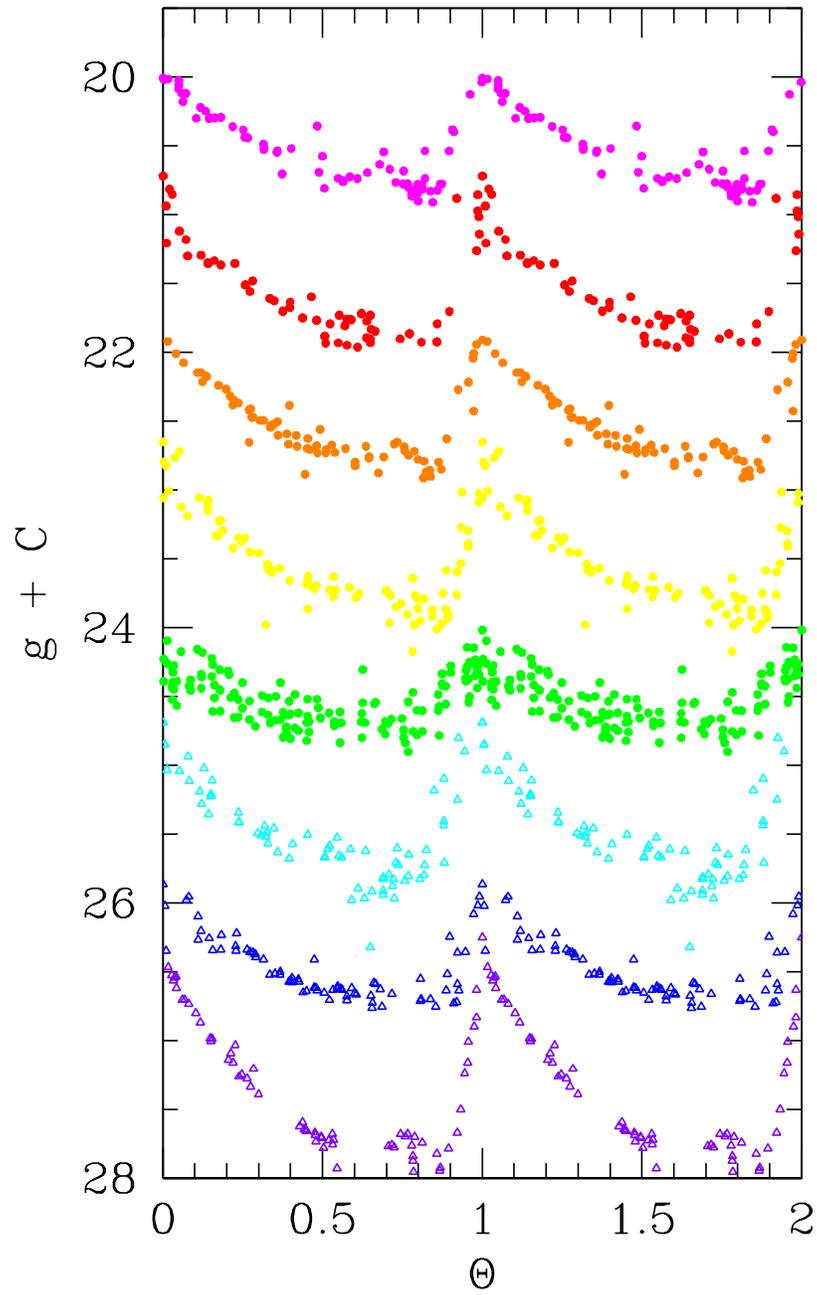}
\caption{Lightcurves for targeted RR Lyrae stars in Stripe-82.  Lightcurves are ordered as they appear in Table 1.  The top five are part of the common-velocity structure near $-75\kms$.}
\label{fig:lightcurves}
\end{figure*}

Obtaining a single-epoch spectrum for an RR Lyrae star yields only a
velocity of the stellar photosphere, which for these pulsating
variables can deviate from the systemic velocity by several 10s of
$\rm km\,s^{-1}$.  However, the dense temporal sampling of Stripe-82
enables us to obtain accurate ephemerides by phase folding the $\sim70$ epochs in the SDSS photometric archive.  We can then correct
velocity measurements made at known phase to the barycentric
velocities for these stars (Joy 1938).  We extracted the lightcurves
from the SDSS archive and determined their periods based on a variant
of the ``Phase Dispersion Minimization'' technique
\citep{stellingwerf78}. We first identified potential bad data points
by performing a regression on $g$ vs.\ $r$ flux, and flagging
$2.5\,\sigma$ outliers (as determined from the scatter -- not the
formal errors).  While the colors of RR Lyrae stars change during
their pulsation cycle, the $2.5\,\sigma$ criterion allows an adequate
range for normal color variation.  We then removed near-achromatic
points that were substantially fainter than the remaining points,
which are probably due to some joint photometric anomaly, but in any
case would not fit any RR Lyrae-like light curve.  Finally, we varied
the period and minimized the sum of the squares of the photometric
differences between successive points.  Our derived $g$-band
lightcurves for our 8 target objects are shown in
Figure~\ref{fig:lightcurves}.  In all but one case, periods derived
from the $g$-band data were nearly identical to periods derived from
the $r$-band data.  Based on these differences, we estimate the period error
to be $\sigma(P)= 3\times 10^{-6}\,$day.  Interestingly, our derived periods are
surprisingly similar, with a dispersion of only $0.026$ days. The
objects are all RRab type variables and with median brightnesses of
about g~$\sim$~20.5 these stars are at a distance of about 85~kpc.  We have also estimated metallicities using the periods and amplitudes for the 8 stars using equation (6) of \citet{sandage04}, assuming that the $V$-band amplitude is equal to the $g$-band amplitude. The metallicities range from [Fe/H]$ \sim$ -1.3 to -2.1. While these estimates are necessarily crude, we
can conclude that all of the 8 stars are relatively metal-poor.

\subsection{Spectroscopic Observations}
The spectroscopic observations were obtained using the Magellan Echellette
Spectrograph (MagE\footnote{http://www.lco.cl/telescopes-information/magellan/instruments/mage/}; Marshall et al. 2008)
mounted on the 6.5-m Clay Telescope at Las Campanas Observatory.  In
all cases two equal-length exposures bracketed an observation of a
ThAr lamp. The data were reduced using a pipeline written by D. Kelson
following Kelson (2003). Post-extraction processing of the spectra was
done with the IRAF\footnote{ IRAF is distributed by the National
  Optical Astronomy Observatories, which are operated by the
  Association of Universities for Research in Astronomy, Inc., under a
  cooperative agreement with the NSF.} {\sc ECHELLE} package.  A 1-arcsec
slit was used, resulting in a resolution of $\sim$4100, and the
signal-to-noise ratio of the final spectra ranged from 7 to 16 per 0.36\AA\
pixel at 4700\AA.  Exposure times varied from 3600s to 4000s depending on the observing conditions.  Velocities were measured with the IRAF FXCOR routine using a
MagE observation of the blue metal-poor star CS22874-009 ($V_{helio}$ =
-36.6 $\kms$, \citealt{ps00}) as the template. The cross
correlations were made on the wavelength interval 4000\AA\ -- 5600\AA\ with
the hydrogen lines masked out.

We adopt the time-averaged velocity of the pulsation curve as the
center-of-mass velocity of the star.  Integration of detailed velocity
curves of the RRa variables WY~Ant, XZ~Aps, DT~Hya, and RV~Oct
(Preston 2009, in preparation) shows that the pulsation velocity is
equal to the star's time-average velocity at phase 0.37, reckoned relative to maximum light. Our
observations were all made as close to this phase as possible, and
velocity corrections were applied adopting $k=92.7$ $\kms$ per unit phase
for the mean slope of the pulsation velocity curve at phase 0.37
for these four stars.

A journal of the observations is presented in Table 1. Column one lists
the heliocentric Julian date at mid-observation followed by the RA and Dec
of the star, the median $g$-band magnitude, the adopted pulsation period, the
observed heliocentric velocity and the error as estimated from the FXCOR
measurement, the number of photometric datapoints in the lightcurve, the heliocentric Julian date at maximum light (zero phase), the pulsation phase at mid-observation, the velocity correction, and the final adopted heliocentric velocity.

\section{Results}

\begin{figure*}
\plotone{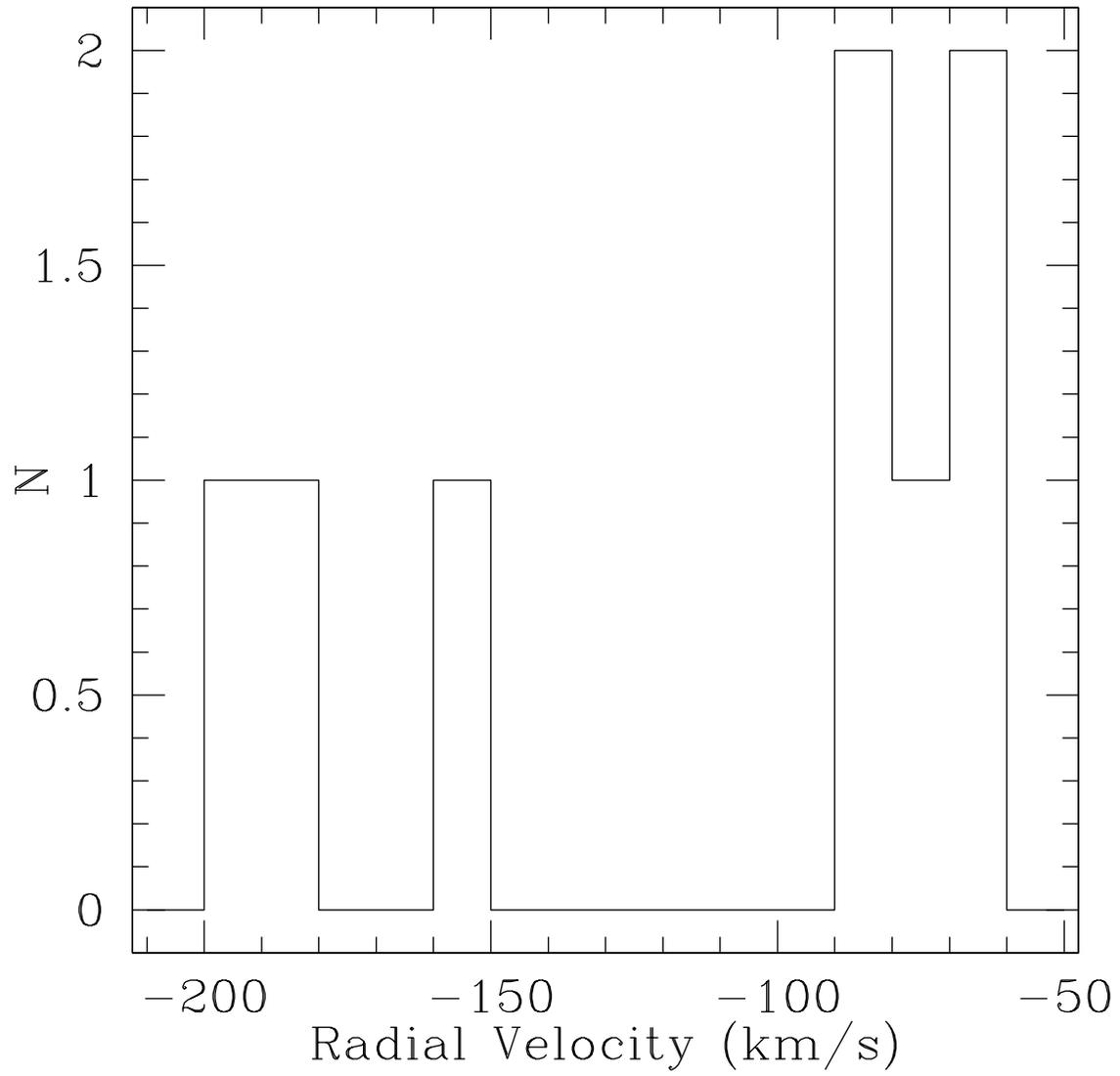}
\caption{Histogram of heliocentric radial velocities for 8 RR Lyrae stars in the Pisces Overdensity.  The concentration at $-75\kms$ is obvious.  The clump at $\sim -190 \kms$ is also of note.}
\label{fig:velhist}
\end{figure*}

\begin{figure*}
\plotone{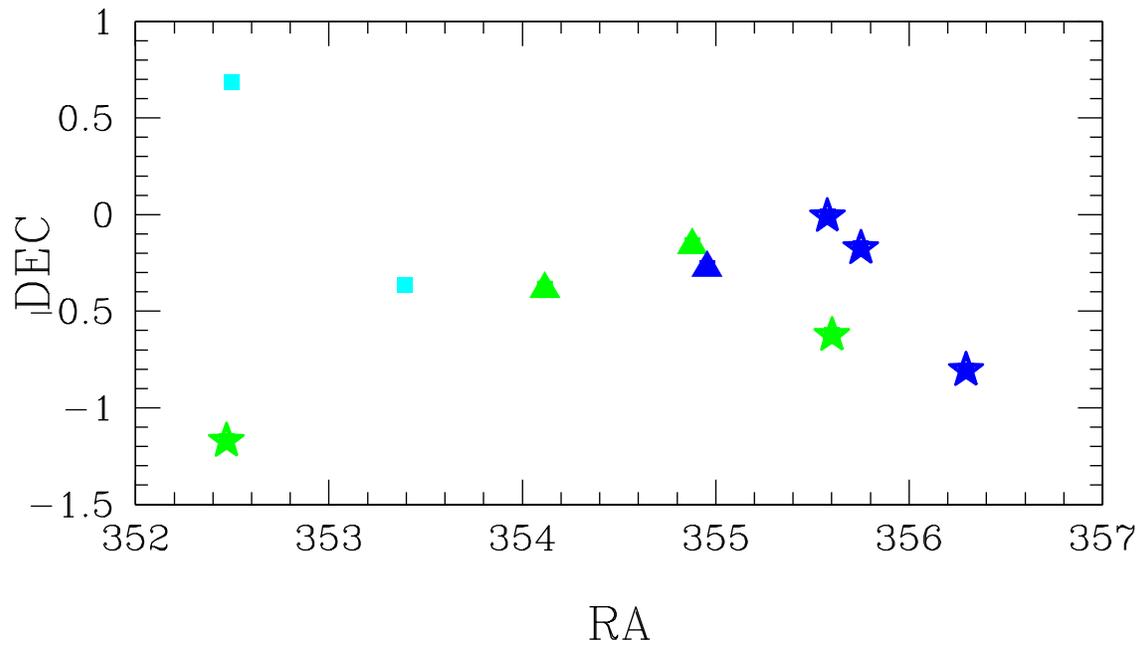}
\caption{RR Lyrae targets for this program.  Green, blue, cyan symbols correspond to median magnitude ranges of $20.5 \leq g_{med} < 20.6$, $20.6 \leq g_{med} < 20.7$, $g_{med}>20.7$.  Star symbols correspond to targets with velocity greater than -90$\kms$ and triangles are all other targets for which we have obtained a radial velocity.  Squares show objects for which we do not yet have spectra.}
\label{fig:targets}
\end{figure*}

We were able to obtain accurate radial velocity measurements for 8 RR Lyrae stars associated within the apparent Pisces Overdensity.   We show in Figure~\ref{fig:velhist} a histogram of the heliocentric radial velocities measured for our targets.  There is a clear velocity peak, containing 5 of the 8 objects, at approximately $-75\kms$.  The other 3 stars also have an interesting velocity distribution to which we return later.    Figure~\ref{fig:targets} shows our targets coded by velocity and magnitude.  We also show in this figure other RR Lyrae stars identified in the field \citep{sesar09} fainter than $V_0=20$.  The spatial concentration at (RA,DEC)$\sim (355^{\circ},-0.3^{\circ})$ is suggestive of a coherent structure.  However, our radial velocity measurements reveal a more nuanced picture.  One of these objects (cyan square) has an apparent magnitude that puts it farther out in the halo than the rest of the concentration.  The three central stars (marked as triangles in the figure) within this concentration have measured radial velocities\footnote{The single object with differing $g$-band and $r$-band period has a barycentric velocity of $-197.7\kms$ if the $g$-band period is adopted and $-182.7\kms$ if the $r$-band period is adopted.} $-198\,{\rm km\,s^{-1}}< v_r < -155\,{\rm km\,s^{-1}}$ .  Four stars in the central concentration and a fifth further away, however, have radial velocities that lie within the main velocity peak.  We therefore seem to have two co-spatial velocity structures, one robustly identified in phase-space and one that is suggestive but too sparsely sampled at present to warrant detailed analysis.  While few data exist at these distances in the stellar halo, a Besan\c{c}on model \citep{robin03} of the heliocentric velocity distribution at this position in the sky suggests that the expected velocity distribution of the smooth stellar halo is centered at $\sim -120\kms$ with a dispersion of $\sim 90\kms$.  In Section \ref{sec:discuss}, we focus exclusively on the main velocity peak.

\section{Discussion }
\label{sec:discuss}

\subsection{Is the Velocity Group Physically Associated?}

In order to determine the random chance that a grouping of 5 out of 8 stars
could have velocities within 20$\kms$ of one another, we performed a
Monte Carlo test by random samplings from a Gaussian
distribution with dispersion of $90\kms$.  We find this probability to
be less than 0.6\%.  This is a fair test (and not an a posteriori
justification of a curious velocity structure found serendipitously)
because this corresponds to exactly the velocity signature we were
looking for when we undertook the observations.  The group is
therefore physically associated at high confidence in a fully Bayesian
sense.

\subsection{Bound or Unbound?}
When looking at the spatial and kinematic information, the first question one must ask is whether these stars are part of a bound or unbound system, i.e., is this an intruding galaxy, or the extended debris of a galaxy/star cluster?  While information on more stars (e.g. giants and horizontal branch stars), would be important to address this question fully, some simple estimates are useful in providing guidance.  

If the system is assumed to be bound, we can estimate the mass
using a virial estimator (e.g., \citealt{heisler85,gould93}):
\begin{equation}
M_{VT}=\frac{3\pi }{2G} \frac{\sigma^2}{\langle 1/R_{\perp,ij}\rangle_{i\not= j}}  
\label{eqn:one}
\end{equation}
where $\sigma$ is the velocity dispersion of the system, $R_{\perp,ij}$ is the projected distance between each two stars.  In order to measure the velocity dispersion we must account for the errors in our velocities.  There are two main (identifiable) sources of velocity error.  The primary source of uncertainty is the error due to the signal-to-noise ratio, which is reported in Table 1.  A secondary source of uncertainty comes from our imprecise knowledge of the zero phase point due to the finite sampling of the Stripe-82 data.  This error is $k/(\sqrt{2}N)=65\kms/N$ where $N$ is the number of data points in each light curve.  For our data, this has a maximum value of $1\kms$ and therefore does not affect our dispersions significantly.  The phase errors induced by the period errors are similar.

Assuming all 5 stars in the main velocity structure are part of the same physical structure, we obtain a velocity dispersion of $6.8^{+3.9}_{-2.6}\kms$ which yields a mass of $1.4^{+1.7}_{-0.7}\times10^{8}\msun$.  A more conservative possibility is that only the ``clump" of four stars near (RA,Dec) = $(356^\circ,-0.3^\circ)$ is part of a bound structure, from which the star at $\sim 352^\circ$ either has been tidally stripped or is not physically associated with the main structure.  In this case, Equation~(\ref{eqn:one}) yields $8.5^{+14.3}_{-5.4}\times 10^7\,M_\odot$, whose $1\,\sigma$ range is not far from the typical value of $1\times 10^7\,M_\odot$ for the mass interior to 300 pc found by \citet{strigari08} for dwarfs with a wide range of luminosities.  Moreover, the apparent $0.5^\circ$ radius of the concentration corresponds to about 750 pc, rather than 300 pc, and this difference could account for the level of discrepancy in mass.  Hence, this structure is possibly virialized, and if so represents a new satellite of the Milky Way.  Given the small number of confirmed stars and the large physical size, however, we cannot at this point rule out the possibility that it is being (or has been) tidally disrupted.

If the apparent structure is physical (and not a chance superposition)
then whether it is bound or not, the presence of 4 RR Lyrae stars
within a $\sim 0.5\,\rm deg^2$ rectangle implies that there
must be a concentration of other stars within the system.
By the fuel consumption theorem, $S_{RR}$, the RR Lyrae specific frequency (the number per $V$-band luminosity, normalized to $M_{V,\rm norm}=-7.5$; \citealt{suntzeff91}) can be cast in terms of the fraction of He burning that takes place within RR Lyrae stars, $\eta_{RR}$:  
%$$
\begin{equation}
S_{RR} = 
{m_{\rm He}-m_{\rm C}/3\over 4 m_{\rm H} - m_{\rm He}}\,
{\eta_{RR}\over 1-\eta_{\rm He}}\,
10^{0.4\Delta M_{\rm bol}} = 360\eta_{RR}
\label{eqn:srr}
\end{equation}
%$$
where $m_{\rm X}$ is the atomic mass of $X$, $\eta_{\rm He}\sim 25\%$
is the initial abundance of He,
$\Delta M_{\rm bol}=M_{V,RR}-M_{V,\rm norm}+\Delta {\rm BC}$,
$M_{V,RR}=0.6$, and $\Delta{\rm BC}=0.45$ is the difference
in bolometric corrections between RR Lyrae stars and typical
giant stars.  The highest observed value is $S_{RR}=158$
\citep{siegel01}, i.e., $\eta_{RR} = 0.44$.  The surface brightness of the 
Pisces concentration may be expressed as
$V = (34.0 + 2.5\log \eta_{RR})\,{\rm mag\,arcsec^{-2}}$, and
hence to be ``recognizable'' as a $V=30\,{\rm mag\,arcsec^{-2}}$
overdensity would require $\eta_{RR} < 0.025$ or
$S_{RR}< 9$.  Such systems do exist \citep{harris96}, but they are far from universal.  We cannot therefore, probe whether the Pisces
structure is a physical (or chance) association based on the presence
(or absence) of an already known photometric counterpart in giants.
Deeper imaging data of this region as well as further spectroscopic
follow-up of giant and horizontal branch stars in this region of
the sky is necessary to confront this hypothesis.

\subsection{Unbound Case: Galaxy or Globular Cluster?}
If the object is an unbound stream, as opposed to a bound system, we must ask whether this represents the disrupted remnant of a globular cluster system or the remnant of a satellite galaxy.  Robust characterization of streams and disrupting galaxies, both theoretically and observationally, is a challenge.  The predicted phase-space morphology depends on a variety of initial conditions including initial mass, M/L ratio, stellar density profile and orbit parameters, and the observations are demanding and not always conducive to multiplexing due to the low density of targets.  A thin physical extent transverse to the direction of motion on the sky, uniform stellar color-magnitude diagram and narrow velocity dispersion would typically indicate a globular cluster stream as opposed to a galaxy stream.  The large width of our structure suggests that if it is disrupting, it is a satellite galaxy as opposed to a globular cluster.  The velocity dispersion is comparable to what is observed within the Sagittarius stream \citep{majewski04} or within the Anticenter Stream \citep{grillmair08}. 

If the Pisces Overdensity is indeed a disrupted satellite it is interesting that located near the central core of this structure are three stars with similar velocities but offset from the main structure by nearly 100$\kms$.  As such large velocity offsets are not plausible simply by disruption effects alone, these stars are either a separate, perhaps bound, structure or they are random interlopers.  The latter scenario is of little interest.  In the former scenario it is intriguing, but perhaps not surprising, that we observe two overlapping streams at this position in the halo.  More locally, the Sagittarius system is so extensive and has made a sufficiently large number of orbits that many objects have been found within it (e.g. Segue 1) that can only be disentangled using velocity and metallicity information.  Farther out in the Galaxy, Bell et al. (2008) have determined that at least 50\% of the stellar density is in a clumpy form.  Detailed comparison with cosmologically-motivated models for the spaghetti-like nature of the halo would be useful in determining whether such self-overlapping systems are rare or commonplace in the Milky Way at these distances.

\section{Conclusions}
We present spectroscopic confirmation of the photometric overdensity observed in RR Lyrae stars toward the constellation Pisces.  We suggest that this system is a dwarf galaxy -- possibly in the process of disruption, possibly already disrupted, or possibly bound with very low surface brightness.

The nature of the stellar halo at distances of $\sim 100\kpc$ is still relatively uncharted territory.  Using RR Lyrae stars to explore this regime of phase space is a novel and exciting way forward.  Large photometric datasets such as those provided by SDSS and similar experiments in conjunction with large ground-based telescopes to obtain spectra should yield important insight into the nature of the Milky Way's perhaps troubled, perhaps tranquil past.

\acknowledgements 

We thank the participants of the KITP program ``Building the Milky
Way'' where this project was hatched.  A.G. was supported by NSF grant
AST-0757888.  I.B.T. was supported by NSF grant AST-0507325.  This
research was supported in part by the National Science Foundation
under Grant No. PHY05-51164.  JAK thanks Andy McWilliam, John Mulchaey, James
Buckwalter, and Luis Saenz for stimulating discussions.

\begin{deluxetable*}{ccccccccccc}
\tabletypesize{\footnotesize}
\tablewidth{0pt}
\tablenum{1}
\tablecolumns{11}
\tablecaption{RR Lyrae Targets}
\tablehead{
\colhead{HJD}&
\colhead{RA}&
\colhead{DEC}&
\colhead{$g_{med}$}& 
\colhead{$P_g$}& 
\colhead{$v_{r,obs}$}&
\colhead{$N_{phot}$}& 
\colhead{HJD$_{\Theta,0}$}&
\colhead{$\Theta_{obs}$}& 
\colhead{$\Delta\,v$}& 
\colhead{$v_{r, bar}$}\\
}
\startdata
5043.8323 & 352.46991 & -1.17125 & 20.545 &  0.5973118 &  -62.8 (4.8) & 65 & 3352.5903 & 0.422  & -4.8    &  -67.6 \\
5042.9037 & 356.29469 & -0.80489 & 20.655 &  0.5955986 &  -73.0 (3.8) & 84 & 4008.7571 & 0.315  &  5.1    &  -67.9   \\
5012.8432 & 355.75079 & -0.17316 & 20.603 &  0.5938712 &  -65.1 (3.3) & 74 & 3270.7662 & 0.425  & -5.1    &  -70.2  \\
5013.9018 & 355.57762 & -0.00838 & 20.628 &  0.5949283 &  -81.9 (3.6) & 80 & 2911.7877 & 0.388  & -1.7    &  -83.6   \\
5044.8612 & 355.60088 & -0.62351 & 20.524 &  0.6314823 &  -83.5 (4.9) & 142 & 3635.7693 & 0.404  & -3.2    &  -86.7   \\ 
5042.8593 & 354.11676 & -0.38425 & 20.569 &  0.6016641 & -156.2 (6.1) & 70 & 4418.7158 & 0.362  &  0.7    & -155.5   \\
5014.8779 & 354.87899 & -0.15772 & 20.569 &  0.5902998 & -191.6 (4.3) & 74 & 3668.7372 & 0.436  & -6.1    & -197.7  \\
5012.8932 & 354.95545 & -0.27631 & 20.607 &  0.5310617 & -191.7 (4.2) & 66 & 3996.7940 & 0.335  &  3.2    & -188.5  \\
\enddata
\tablecomments{Periods derived in $g$ and $r$-bands were identical for all but 1 star.  In that case the velocity changes from -197.7 $\kms$ to -182.7$\kms$. Quoted velocities are heliocentric and values in parenthesis are the estimated observational errors.}
\label{tab:tab_rr}
\end{deluxetable*}

\end{document}